\documentclass[journal]{IEEEtran}
\usepackage{cite}
\usepackage{graphicx}
\usepackage{subfig}
\usepackage{amssymb,amsmath} 
\usepackage{multirow}

\begin{document}

\title{Validation of Compton Scattering Monte Carlo Simulation Models}

\author{Georg~Weidenspointner, 
        Matej~Bati\v{c}, 
        Steffen~Hauf, 
        Gabriela~Hoff, 
        Markus~Kuster,
        Maria~Grazia~Pia, 
        and~Paolo~Saracco  
\thanks{Manuscript received \today}
\thanks{This work was supported by XFEL GmbH in the framework of the
  DSSC project.}
\thanks{G.~Weidenspointner is with Max-Planck-Institut f\"ur
  extraterrestrische Physik, Giessenbachstrasse, 85741 Garching,
  Germany (e-mail: ggw@hll.mpg.de).}
\thanks{M.~Bati\v{c} was with INFN Sezione di Genova, Genova, Italy
  (e-mail: Batic.Matej@gmail.com); he is now with Sinergise, 1000
  Ljubljana, Slovenia.}
\thanks{S.~Hauf and M.~Kuster are with European X-ray Free Electron
  Laser facility GmbH, Albert-Einstein-Ring 19, 22761 Hamburg, Germany
  (e-mail: steffen.hauf@xfel.eu,
  markus.kuster@xfel.eu).}
\thanks{G.~Hoff is with Pontificia Universidade Catolica do Rio Grande
  do Sul, Rio Grande do Sul, Brazil (e-mail: ghoff@gmail.com).}
\thanks{M.G.~Pia and P.~Saracco are with INFN Sezione di Genova, Via
  Dodecaneso 33, 16146 Genova, Italy (e-mail:
  MariaGraziaPia@ge.infn.it, PaoloSaracco@ge.infn.it}
}  

\maketitle
\thispagestyle{empty}


\begin{abstract}

  Several models for the Monte Carlo simulation of Compton scattering
  on electrons are quantitatively evaluated with respect to a large
  collection of experimental data retrieved from the literature. 
%
%
  Some of these models are currently implemented in general purpose
  Monte Carlo systems; some have been implemented and evaluated for
  possible use in Monte Carlo particle transport for the first time in
  this study.
  Here we present first and preliminary results concerning total and
  differential Compton scattering cross sections.





\end{abstract}



%


\section{Introduction}

%
%
%
%
\IEEEPARstart{C}{ompton} scattering is of fundamental importance for
hard X-ray and soft $\gamma$-ray imaging and polarimetry systems in a
wide range of applications in nuclear science, astronomy, medical
science, radiation safety, or homeland security. It is the dominant
interaction process of photons with matter at hard X-ray and soft
$\gamma$-ray energies. The inelastic scattering of photons on
electrons was discovered in 1923 by A.H.~Compton \cite{Compton1923};
for stationary and free electrons, the relativistic theory of this
process was presented in 1929 by O.~Klein and Y.~Nishina
\cite{KleinNishina1929}. When considering the scattering of photons in
matter, electrons are neither stationary nor free, rather they are
bound to atomic nuclei and possess non-zero orbital momentum. For
photon energies below a few 100~keV, the effect of electron binding is
not negligible for inelastic scattering and can be described by a
so-called incoherent scattering function (see e.g.\
\cite{Ribberfors1975}).

Compton scattering is simulated by all general purpose Monte Carlo
systems that model the transport of photons and other particles in
matter (e.g.\ EGS~\cite{EGS}, FLUKA~\cite{FLUKA},
Geant4\cite{geant403,geant406}, ITS\cite{ITS}, MCNP~\cite{MCNP}, and
Penelope\cite{Penelope}).  Nevertheless, an objective, quantitative
evaluation of the physical accuracy of Compton scattering simulation
models is not yet documented in the literature. The validation of
Compton scattering models implies their comparison with experimental
data \cite{Trucano2006}. The accuracy of simulation models can then be
objectively quantified based on rigorous statistical analysis.

The validation of Compton scattering models presented here is part of
a larger effort to validate photon interaction simulation models for
general purpose Monte Carlo systems. An overview can be
found in \cite{Batic2013}, validation results for photon elastic
scattering have been reported in \cite{Batic2012}, preliminary
results for photo-ionization and electron-positron pair production
were presented at this conference (see \cite{Basaglia2013},
\cite{Begalli2013}). 

In this article, we present first and preliminary validation results
concerning a variety of models for total and differential Compton
scattering cross sections, with emphasis on those implemented in
Geant4, with respect to a large collection of experimental data
retrieved from the literature.

\section{Compton Scattering in Geant4 9.6}
\label{sec:models}

\begin{figure*}[!t]
\centerline{\includegraphics[width=7in]{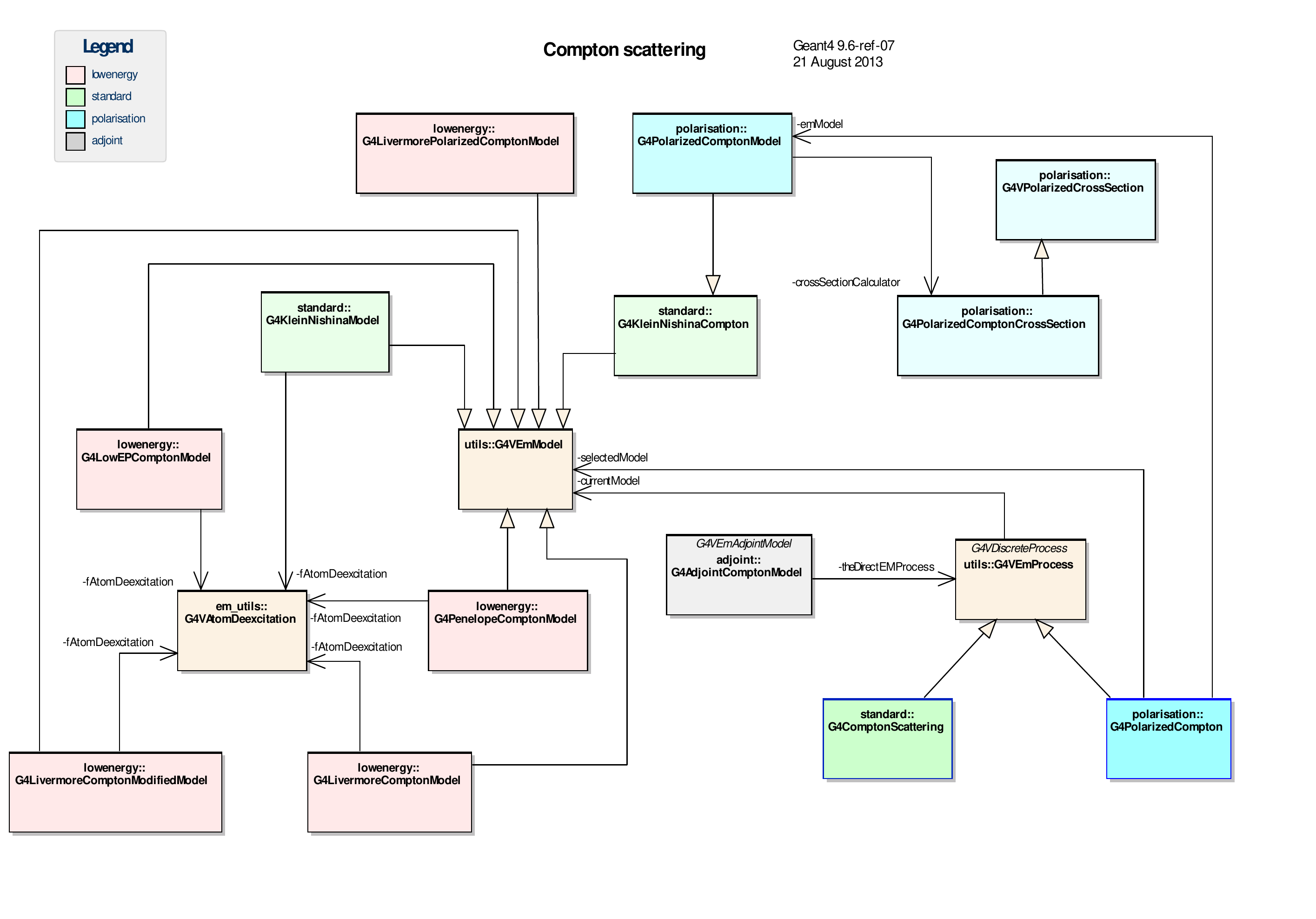}}
\caption{A UML class diagram of the Compton scattering models
  implemented in Geant4 9.6. The color coding categorizes the models
  into standard, low-energy and polarized physics simulation as well
  as reverse Monte Carlo via the adjoint physics implementation.}
\label{fig:uml_class_geant4}
\end{figure*}

Geant4, as of version 9.6, currently implements 10 Compton scattering
models, and physics processes that use these models. An overview of
the models and their dependencies is given in the UML (Unified
Modeling Language) class diagram shown in
Fig.~\ref{fig:uml_class_geant4}. As is indicated in the class diagram,
these models can be subdivided into four general physics scenario
categories: standard, low-energy, polarized and adjoint. The following
discussion, which summarizes observations on the code with respect
to good design practices as given e.g. by Fowler\cite{fowler},
considers only the first two models, as they constitute the most
common application scenarios.

The Geant4 ``standard'' electromagnetic package encompasses two
classes implementing Compton scattering models based on the
Klein-Nishina\cite{KleinNishina1929} approach:
{\it G4KleinNishinaCompton} and {\it G4KleinNishinaModel}. These
models are selectable for usage by the {\it G4ComptonScattering}
process, which is derived from {\it G4VEmProcess}, which in turn is a
concrete implementation of the {\it G4VDiscreteProcess} base class
encompassed in the Geant4 kernel.  Additionally, a {\it
  G4HeatedKleinNishinaCompton} model is present, which can be used to
simulate comptonization in a hot plasma. It does not appear to be
documented in the Geant4 physics reference manual~\cite{geant:physics}
and will not be further discussed here.

The initially mentioned two models are in fact very similar, in that
they duplicate code for an empirically parameterized cross section
calculation~\cite{geant:physics} derived from theoretical tabulations
by~\cite{hubbell,storm_israel}. They differ in the final state
generation. {\it G4KleinNishinaModel} handles the atomic relaxation
following the emission of an electron from the target atom
by delegating it to {\it G4AtomicDeexcitation}, while {\it
  G4KleinNishinaCompton} does not take into account atomic relaxation.
The code duplication present in the cross section calculation is
unnecessary and should be avoided \cite{fowler}.


For applications which require the simulation of Compton scattering at
lower energies than the 10~keV limit documented for the ``standard''
models ~\cite{geant:physics}, the Livermore- and Penelope-based
Compton scattering models can be used.

In Geant4 9.6 these models consist of {\it
G4LivermoreComptonModel}, {\it G4LivermoreCompton-ModifiedModel}, {\it
G4LowEPComptonModel} and {\it G4Penelope-ComptonModel}. Here the first
three models implement a cross section calculation based on the EPDL data
library~\cite{epdl}, while the latter uses an approach reengineered 
from the Penelope code~\cite{penelope}.

In their final state generation the Livermore-based models describe 
the scattered photon distribution using a scattering function as given
by Cullen~\cite{cullen_sf}. {\it G4LivermoreComptonModel} and
{\it G4LivermoreComptonModifiedModel} differ in the modeling of the
scattered electron, whereas {\it G4LowEPComptonModel} adds a
fully relativistic treatment to the final state generation using the
Relativistic Impulse Approximation (RIA)~\cite{RIA, geant:physics}.
Across all Livermore-based models, code duplication is again a major
issue. The code has been fully duplicated for the
cross section calculation, and partially duplicated for the final
state generation.



\section{Refactored software}
\label{sec:refactor}

\begin{figure}
\centerline{\includegraphics[width=8.5cm]{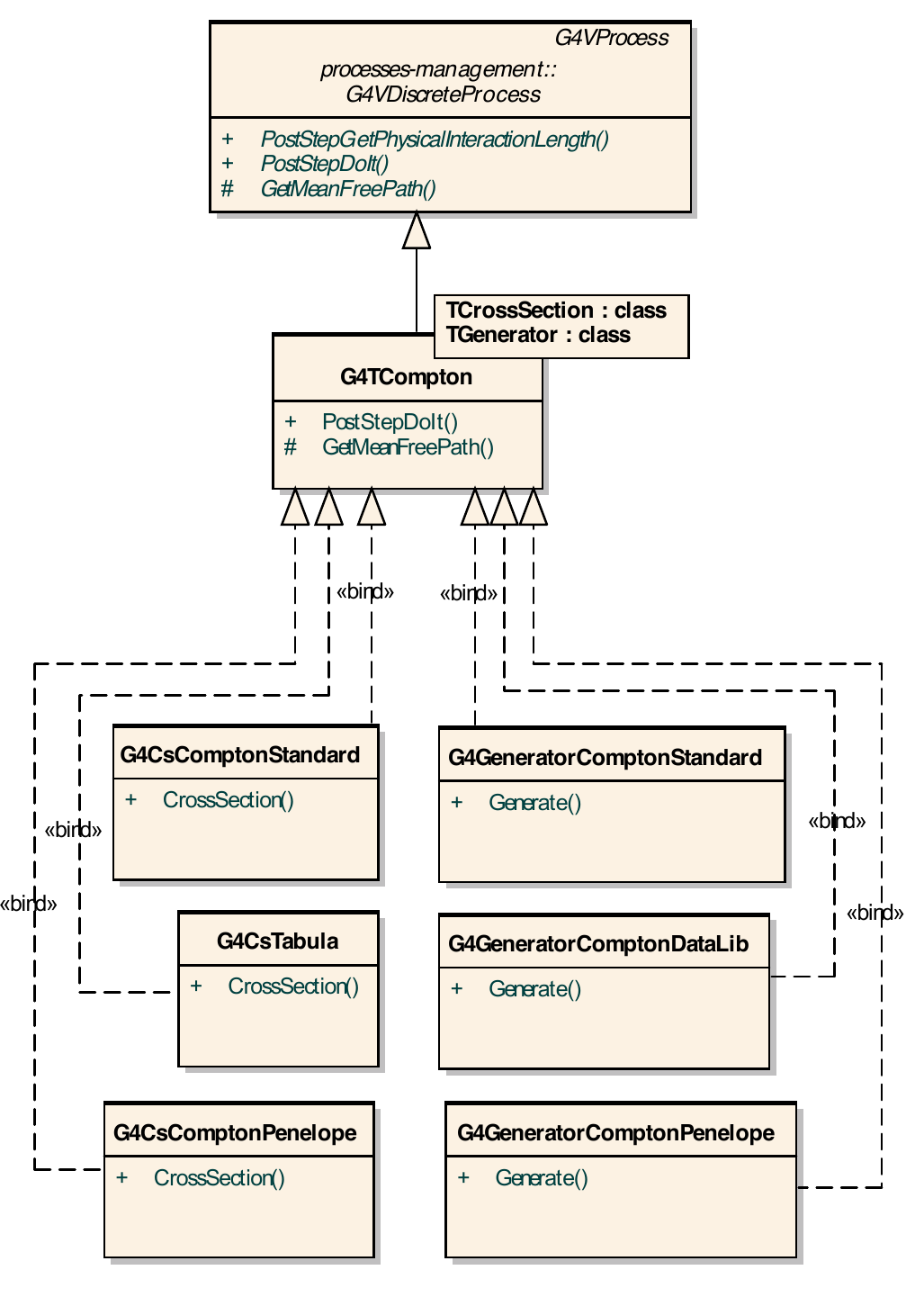}}
\caption{
A UML class diagram of the refactored policy-based design for Geant4
Compton scattering. Policy classes for final state and cross section
generators are bound to the G4TCompton host template class.}
\label{fig:uml_refactor}

\end{figure}

The software design has been refactored based on a sharp domain
decomposition, which identified total cross section calculation and
final state generation as two distinct entities of the problem domain.
A policy-based class design \cite{alexandrescu} has been adopted: it
ensures flexibility at endowing the Compton scattering process with
multiple behaviours based on a variety of alternative modeling
approaches, while the intrinsic simplicity, restricted
responsibilities, and minimized dependencies of policy classes
facilitate the testing of the software both in the processes of
verification and of validation. The same software design approach has
been successfully adopted in the simulation of photon elastic
scattering \cite{tns_rayleigh}. The physics functionality of policy
classes can be tested by means of simple unit tests, whereas the
design of the Compton models in Geant4 9.6 hinders testing the physics
functionality of the models outside a full simulation environment.
The main features of the refactored software design, which does not
alter physics functionality, are illustrated in the UML class diagram
of Fig.~\ref{fig:uml_refactor}.

In the first development cycle documented in this paper three total
cross section policies were implemented, which correspond to the three
uniquely distinct modeling approaches documented in
Section~\ref{sec:models}: standard (empirical fit to
~\cite{hubbell,storm_israel}), EPDL-based and Penelope-based
(Penelope).

Similarly, three differential cross section policies were implemented,
which correspond to three distinct methods used to sample the
scattered photon angular distribution implemented in Geant4: according
to the Klein-Nishina formula, according to the Klein-Nishina formula
multiplied by a scattering function, and according to the method
reengineered from Penelope.  In addition to the tabulated scattering
function data of EPDL, which are used by Geant4, scattering function
tabulations based on work by Brusa et al.~\cite{brusa}, Biggs et
al.~\cite{biggs}, and Hubbell et al.~\cite{hubbell_sf} were
reformatted to be used by the above mentioned policy class, and
included in the validation of differential cross sections. The
tabulations identified as ``Hubbell'' and as EPDL are based on the
same calculations~\cite{hubbell_sf}, but they differ in the
sin$(\theta/2)$ values at which scattering functions were calculated.

\section{Data Extraction}

\subsection{Experimental Data}

The experimental data were extracted from the literature,
yielding 230 data points for total cross sections and 2612 data points
for differential cross sections for elements ranging from hydrogen
($\mathrm{Z}=1$) to uranium ($\mathrm{Z}=92$).

For this preliminary analysis the data have been extracted into
tabular format and converted to units of barn or barn/sr whenever
necessary and appropriate. Cross sections derived from subtracting
theoretically calculated photoelectric and elastic scattering
contributions from total photon attenuation coefficient measurements
were excluded from the validation analysis.  The tabular format is
such that data are accessible by incident photon energy as well as
scattering angle. A full screening for and removal of outliers, as
well as a consistency check of experimental uncertainties, has not yet
been performed for this conference contribution, but will be included
in our final analysis.

\subsection{Simulation models}

Total and differential cross sections for the incident photon
energies, scattering angles and target elements covered by the
experimental data were obtained by means of unit tests.

For the validation of total scattering cross sections unit tests were
done for the three distinct policy classes described in
Section~\ref{sec:refactor}: standard (empirical fit to
~\cite{hubbell,storm_israel}) , EPDL-based and Penelope-based
(Penelope).
Similarly, the validation of differential cross sections involved unit
tests associated with the three options mentioned in
Section~\ref{sec:refactor}.  In addition to using scattering functions
tabulated in EPDL, unit tests were performed using the aforementioned
alternative scattering functions.

\section{Validation Strategy}

For the validation the different models have been compared to the
individual experimental data points using a $\mathrm{\chi}^2$
goodness-of-fit test. If the p-value of a given test exceeded a
significance value of $\mathrm{\alpha}=0.01$, i.e.
$\mathrm{p}(\mathrm{\chi}^2)\ge\mathrm{\alpha}$, the test was
classified as passed, i.e.\ the model was considered compatible with
the data. In cases where $\mathrm{p}(\mathrm{\chi}^2)<\mathrm{\alpha}$
the test was considered as failed, i.e. model and data were considered
incompatible. The efficiency of a given cross section model is then
determined by the fraction of test cases which were found to be
compatible with the data with respect to the total number of test
cases for this model:
\begin{equation}
	\mathrm{\epsilon} = \frac{N_{\mathrm{p\ge0.01}}}{N_{\mathrm{total}}}.
\end{equation}

In the case of total cross sections one test case consists of all
available data points. For differential cross sections all data
available for a given energy and scattering angle constitute one test
case. The total efficiency of a model was then calculated as the mean
of the efficiencies obtained from all test cases of this model.

\section{Results}

\begin{table}
  \caption{Comparison of efficiencies of validated differential cross
    section models. Note that Klein-Nishina represents the
    implementation in Geant4 standard physics.} 
\label{tab:results}
\centerline{\begin{tabular}{l|r}
{\bf model} & {\bf efficiency} \\
\hline
\hline
EPDL		  & $0.82\,\pm\,0.02$ \\
Penelope	  & $0.82\,\pm\,0.02$ \\
Klein-Nishina & $0.54\,\pm\,0.03$ \\
Brusa		  & $0.84\,\pm\,0.02$ \\
BrusaF		  & $0.84\,\pm\,0.02$ \\
PenBrusa	  & $0.84\,\pm\,0.02$ \\
PenBrusaF	  & $0.84\,\pm\,0.02$ \\
Biggs		  & $0.84\,\pm\,0.02$ \\
BiggsF		  & $0.85\,\pm\,0.02$ \\
Hubbell		  & $0.82\,\pm\,0.02$ \\
\end{tabular}}
\end{table}

\begin{figure*}[!ht]
\centerline{\subfloat[]{\includegraphics[width=2.5in]{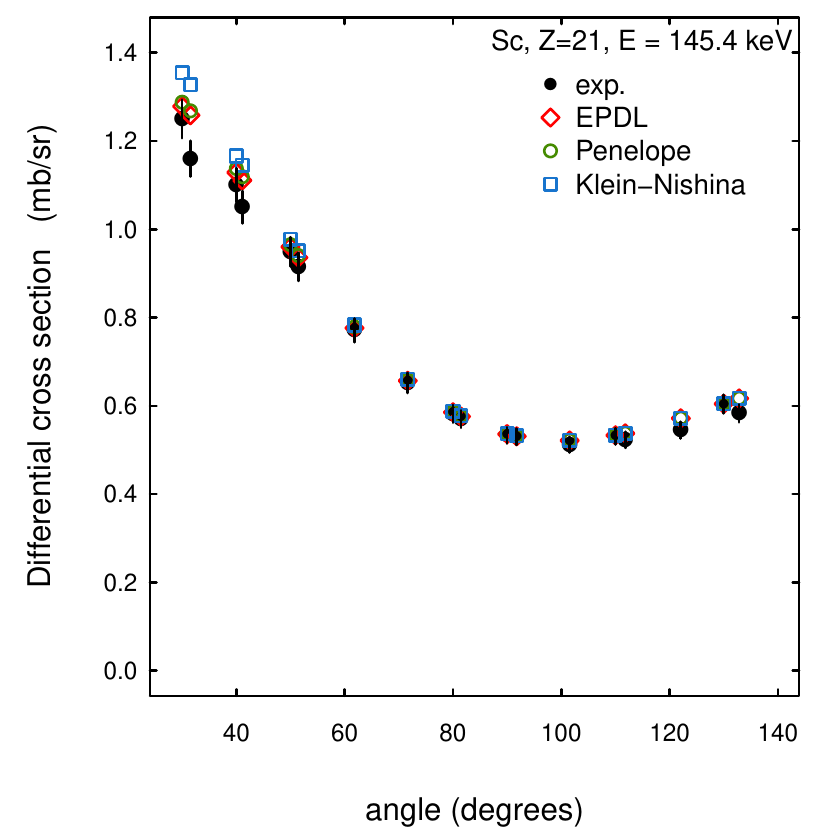}%
\label{res_21}}
\hfil
\subfloat[]{\includegraphics[width=2.5in]{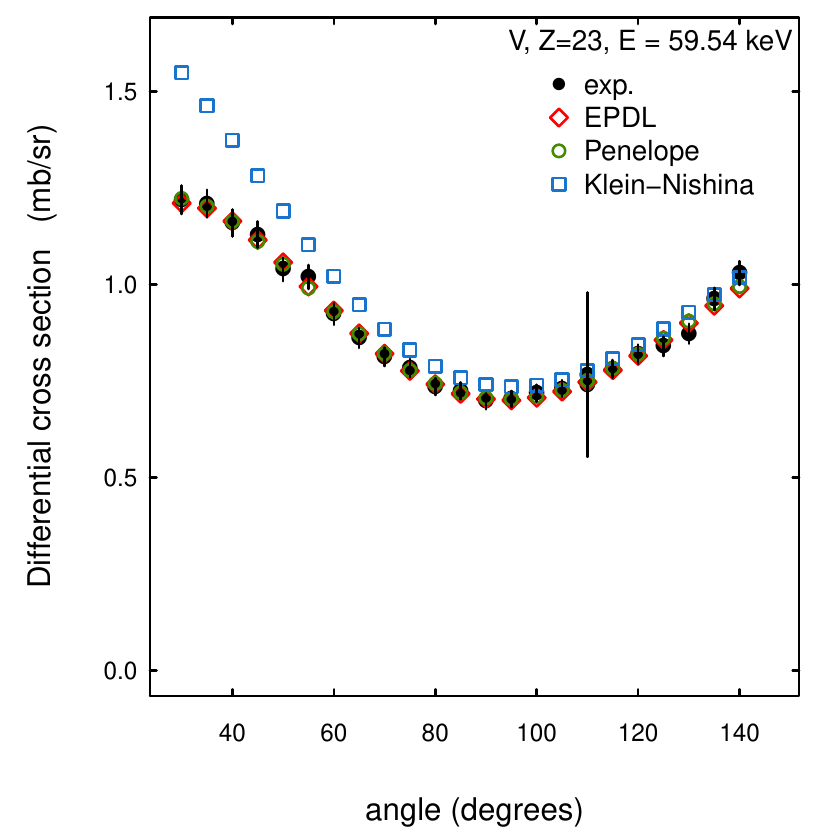}%
\label{res_22}}}
\centerline{\subfloat[]{\includegraphics[width=2.5in]{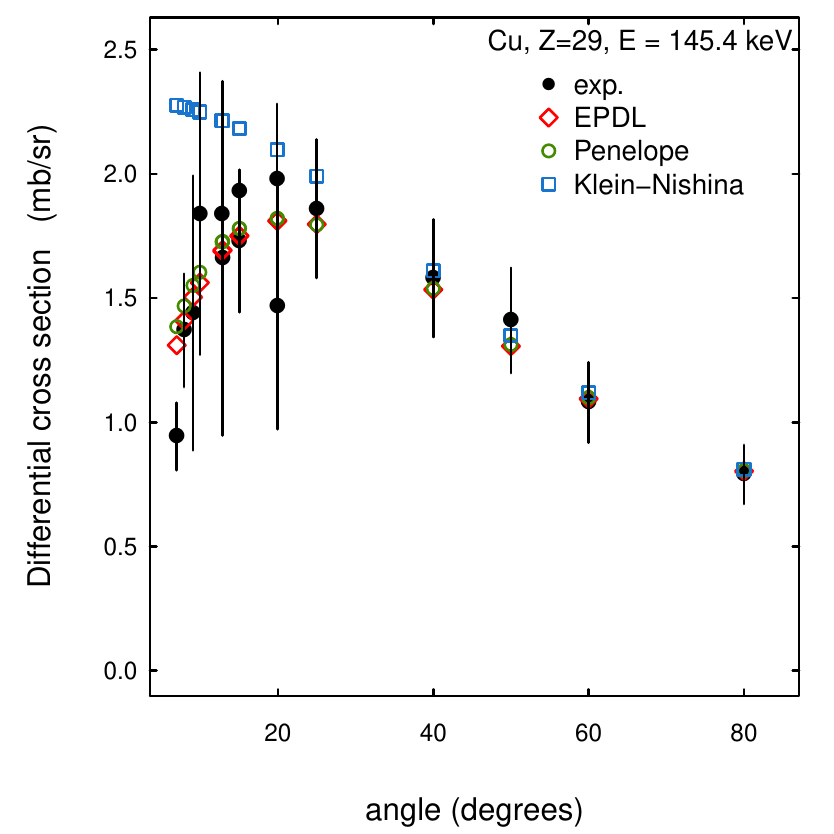}%
\label{res_29}}
\hfil
\subfloat[]{\includegraphics[width=2.5in]{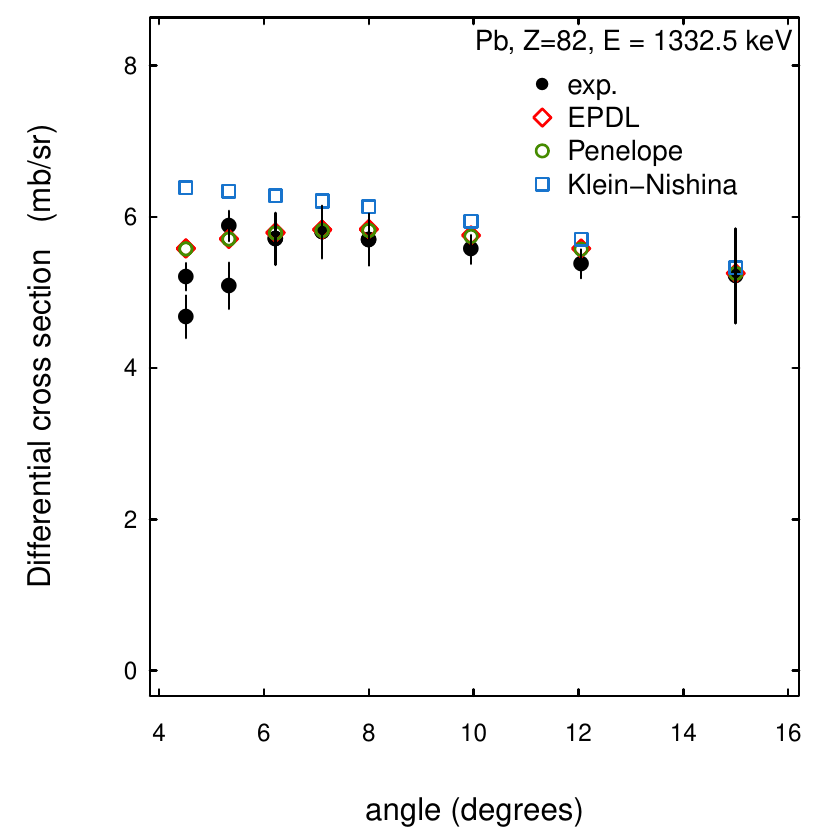}%
\label{res_82}}}
\caption{Validation results for differential cross sections of
  elements Sc (Z = 21), V (Z = 23), Cu (Z = 29) and Pb (Z = 82). The
  Penelope- (green circles) and EPDL-based (red diamonds) models are
  often indistinguishable and generally describe the experimental data
  well. The Klein-Nishina-based model (blue squares) tends to
  overestimate the differential cross section in these examples, in
  particular at low scattering angles. 
%
%
  In some cases, there are noticeable differences between experimental
  data sets.
}
\label{fig_results}
\end{figure*}

Concerning total Compton scattering cross sections our preliminary
analysis shows that all tested models reach an efficiency of $1$ for
all test cases, i.e.\ all models are capable of modeling the
experimental data with a significance of $\mathrm{\alpha}=0.01$.

Table~\ref{tab:results} shows the mean efficiencies achieved by the
different models when compared to experimental differential Compton
scattering cross sections. It is apparent that all models except the
Klein-Nishina model behave almost equally well. This observation is
also qualitatively supported by the exemplary plots of differential
cross sections for different elements and energies shown in
Fig.~\ref{fig_results}. A more detailed follow-up to the preliminary
analysis presented here, which categorizes e.g.\ different energy
ranges and angles, and corrects for outliers in the experimental data,
will likely distinguish more between these models.


\section{Summary and Prospects}

Our first and preliminary results concerning the validation of total
Compton scattering cross sections suggest that all evaluated models
agree equally well with available experimental data. Concerning
differential Compton scattering cross sections, we arrive at a similar
conclusion, with the notable exception of models based on the original
Klein-Nishina theory, as is e.g. the case for the Compton scattering
models implemented in Geant4 standard physics. This was to be
expected, since the original Klein-Nishina theory describes Compton
scattering on free electrons, and therefore by design does not take
into account binding effects for electrons bound to an atom.

Before final conclusions can be reached, we will subject our data base
of experimental results to a more detailed critical appraisal to
identify possible systematic biases or outliers. At the same time, we
will evaluate additional Compton scattering models.

Further aspects of Compton scattering that still await validation are
shell cross sections, Doppler broadening, polarization, and finally
computational efficiency. 

The complete set of results of the validation of Compton scattering
simulation, including additional physics features and modeling options
that are not considered in this paper, will be reported in detail in a
forthcoming publication in a refereed journal.

\end{document}